\documentclass[11pt]{article}

\usepackage{scicite}
\usepackage{graphicx}
\usepackage{times}
\usepackage{amsmath, amssymb, amsfonts, amsthm, fouriernc}
\usepackage{mathtools}
\usepackage[nodisplayskipstretch]{setspace}
\usepackage{comment}
\usepackage{xcolor}
\graphicspath{{figures/}}

\newenvironment{sciabstract}{%
\begin{quote} \bf}
{\end{quote}}

\title{Explosive synchronization in a turbulent reactive flow system}

\author
{Amal Joseph$^{1}\footnote{These authors contributed equally to this work}$, Induja Pavithran$^{2,3\ast}$, R. I. Sujith$^{2,3\dagger}$\\
\\
\normalsize{$^{1}$Department of Mechanical Engineering, College of Engineering, Trivandrum 695016, India}\\
\normalsize{$^{2}$Department of Aerospace Engineering, Indian Institute of Technology Madras, Chennai 600036, India}\\
\normalsize{$^{3}$Center of Excellence for studying Critical Transitions in Complex Systems,} \\ {Indian Institute of Technology Madras, Chennai 600036, India}\\
\\
\normalsize{$^\dagger$To whom correspondence should be addressed; E-mail:  sujith@iitm.ac.in}
}

\date{}

\begin{document} 


\baselineskip24pt


\maketitle


\begin{sciabstract}
 The occurrence of abrupt dynamical transitions in the macroscopic state of a system has received growing attention. We present experimental evidence for abrupt transition via explosive synchronization in a real-world complex system, namely a turbulent reactive flow system. % also known as a thermo-acoustic system. 
 In contrast to the paradigmatic continuous transition to a synchronized state from an initially desynchronized state, the system exhibits a discontinuous synchronization transition with a hysteresis. We consider the fluctuating heat release rate from the turbulent flames at each spatial location as locally coupled oscillators that are coupled to the global acoustic field in the confined system. We analyze the synchronization between these two subsystems during the transition to a state of oscillatory instability and discover that explosive synchronization occurs at the onset of oscillatory instability. Further, we explore the underlying mechanism of interaction between the subsystems and construct a mathematical model of the same.
\end{sciabstract}

%\paragraph{Short title:} Explosive synchronization in complex systems
%\paragraph{Teaser:} Chaotic heat release rate fluctuations from turbulent flamelets synchronize explosively, leading to an abrupt transition to a periodic state.
%Explosive synchronization in spatial dynamics is accompanied by a gradual, continuous synchronization in the mean-field dynamics.
% Explosive synchronization of spatially distributed oscillators is accompanied by a gradual synchronization between the mean-field dynamics.
% A gradual synchronization between the mean-field dynamics of the subsystems leads to an explosive synchronization.
% We experimentally reveal explosive synchronization in a reactive turbulent system and model it using network of oscillators
% We reveal the mechanism of explosive synchronization to an ordered dynamics in a reactive turbulent flow system.

% In setting up this template for *Science* papers, we've used both
% the \section* command and the \paragraph* command for topical
% divisions.  Which you use will of course depend on the type of paper
% you're writing.  Review Articles tend to have displayed headings, for
% which \section* is more appropriate; Research Articles, when they have
% formal topical divisions at all, tend to signal them with bold text
% that runs into the paragraph, for which \paragraph* is the right
% choice.  Either way, use the asterisk (*) modifier, as shown, to

\section*{Introduction}
Complex systems are ubiquitous in the real world, and they play an increasingly important part in our lives. Natural systems such as climate \cite{baede2001climate}, ecosystems \cite{gaucherel2020understanding}, and biological systems \cite{koch1999complexity} and human-made systems such as power networks \cite{nardelli2014models} and transportation systems are a few examples. They consist of several interacting subsystems, and the collective dynamics of such coupled subsystems leading to emergent behaviour \cite{bar2002general} have garnered attention in the last few decades. 
Many complex systems undergo phase transitions as system parameters are varied. Here, the term phase of a system denotes one of its states, characterized by a set of physical properties that can be considered uniform over a macroscopic length scale.
In such scenarios, the system transitions to a coherent state as the coupling strength between the subsystems is increased (generally achieved by varying control parameters in experiments). Such transitions can be described by the variation of an order parameter (a scalar, a vector, or even a tensor that measures the degree of order in the system, the magnitude of which usually ranges from zero to one) as a function of the control parameter. The framework of synchronization theory allows us to describe the coupled behaviour of the components during dynamical transitions in complex systems.

Synchronization is the emergence of collective behavior among coupled systems where inanimate or living systems adjust their own rhythms due to coupling \cite{pikovsky2003synchronization}. The synchronization of coupled oscillators, including those displaying chaotic dynamics, has been studied in various systems by researchers around the globe \cite{pikovsky2003synchronization,boccaletti2002synchronization,rosenblum1996phase,fujisaka1983stability,manoj2018experimental}. 
Following the seminal work of Winfree and Kuramoto \cite{winfree1967biological,kuramoto1984chemical}, the majority of subsequent studies focused on synchronization transitions of second-order, where the synchronized clusters grow gradually, leading to a continuous and reversible transition \cite{kiss2002emerging,chen2002phase,karnatak2009synchronization}. However, switch-like abrupt transitions to synchronization have been studied in networks of oscillators \cite{cualuguaru2020first,leyva2013explosive}, where a sudden spontaneous emergence of a synchronized macroscopic state of the system occurs; this is referred to as explosive synchronization. 

A crucial property of explosive synchronization is the existence of a hysteresis associated with the transition to synchronization \cite{zou2014basin}. Such an irreversible and sudden change in the state of the system is often undesirable in real-world systems. Researchers first revealed explosive synchronization in coupled Kuramoto oscillators, and it was subsequently extended to networks of chaotic units and confirmed by an experiment using electronic circuits \cite{kumar2015experimental}. Later, Kuehn and Bick \cite{kuehn2021universal} mathematically demonstrated that, in nonlinear dynamical systems, the change from continuous to discontinuous transitions is not surprising but rather an expected effect when additional parameters are varied.

Yet, experimental studies of explosive synchronization are limited to laboratory realizations with electronic circuits and chemical oscillators \cite{kumar2015experimental,cualuguaru2020first}. An understanding of the dynamics during explosive synchronization in spatially extended real-world systems has been found lacking. Despite the lacuna in research  towards explaining this phenomenon in real-world complex systems, abrupt transitions are observed everywhere, ranging from cascading failures in power grids \cite{buldyrev2010catastrophic} to epileptic seizures in the brain \cite{adhikari2013localizing,wang2017small}.  There are no known attempts to study explosive synchronization in practical complex systems where controlled experiments can be conducted by varying parameters. 

Turbulent fluid mechanical systems are real-world complex systems where controlled experiments can be performed. 
One such practical system is a combustion-based power-generating system that is used in land-based gas turbine engines, rocket engines, and jet engines. When combustion occurs in an unsteady flow field in a confinement, the perturbations in the flow cause the flame to fluctuate, resulting in the generation of sound waves. These sound waves are reflected back from the boundaries and affect the heat release rate, thus establishing a feedback loop. A positive feedback can result in the growth in the amplitude of acoustic pressure oscillations. These systems are termed thermo-acoustic systems \cite{sujith2021thermoacoustic}.
% Laboratory-scale setups of thermo-acoustic systems have been used to study synchronization transitions \cite{pawar2017thermoacoustic}. 
Thermo-acoustic systems with a turbulent reactive flow are highly susceptible to oscillatory instabilities, resulting in large amplitude self-sustained oscillations. This phenomenon is better known as thermo-acoustic instability. Spontaneous emergence of such high-amplitude oscillations is undesirable and leads to severe vibrations, fatigue, wear and tear, structural damage, increased heat transfer overwhelming the thermal protection system, reduces the lifespan of the combustor, and can even lead to forced shutdown of power-producing gas turbines and mission failures in the case of rockets \cite{lieuwen2005combustion,fisher2009remembering}.

%Oscillatory instabilities in fluid mechanics is a real world application of complex network dynamics. They manifest as high amplitude limit-cycle oscillations in thermo-acoustic, aeroacoustic, aeroelastic systems. Thermo-acoustic systems with a turbulent reactive flow are highly susceptible to oscillatory instabilities, resulting in large-amplitude self-sustained oscillations, better known as thermo-acoustic instability. Spontaneous emergence of such high -amplitude oscillations results in severe vibrations, fatigue, wear and tear, structural failure, and reduced lifespan of the combustor. In addition to this, it leads to forced shutdowns of power-producing gas turbines and mission failures in the case of rockets in extreme cases (Ref).

Thermo-acoustic instability arises due to positive feedback established as a result of the interactions between the subsystems, i.e., the turbulent flow, unsteady heat release from combustion, and the acoustic field in the confinement. A turbulent thermo-acoustic system is a complex system wherein the nonlinear interactions between the subsystems lead to the emergence of self-organized and ordered dynamical states. The onset of thermo-acoustic instability in turbulent reactive flow systems is a spatio-temporal transition from chaos to order \cite{george2018pattern,mondal2017onset}. Pawar et al. \cite{pawar2017thermoacoustic} studied this transition in the framework of synchronization between the acoustic field and the unsteady heat release rate. They described the onset of thermo-acoustic instability as a synchronization transition from a desynchronized state to intermittent synchronization, then to phase synchronization, and finally to generalized synchronization. 
In contrast to such a gradual transition between synchronization states, explosive synchronization would be more dangerous in practical thermo-acoustic systems such as turbulent combustors used in gas turbines and rockets. Even though such a phenomenon can be expected in a spatially extended complex system such as a turbulent fluid mechanical system, there are no experimental studies on the evidence of explosive synchronization in such real-world complex systems. 
In this study, we identify the occurrence of explosive synchronization leading to an abrupt transition to oscillatory instability in a turbulent thermo-acoustic system. 

\section*{Results}
\subsection*{Experimental demonstration}
\begin{figure*}[h!]
\begin{center}
\includegraphics[scale = 0.82] {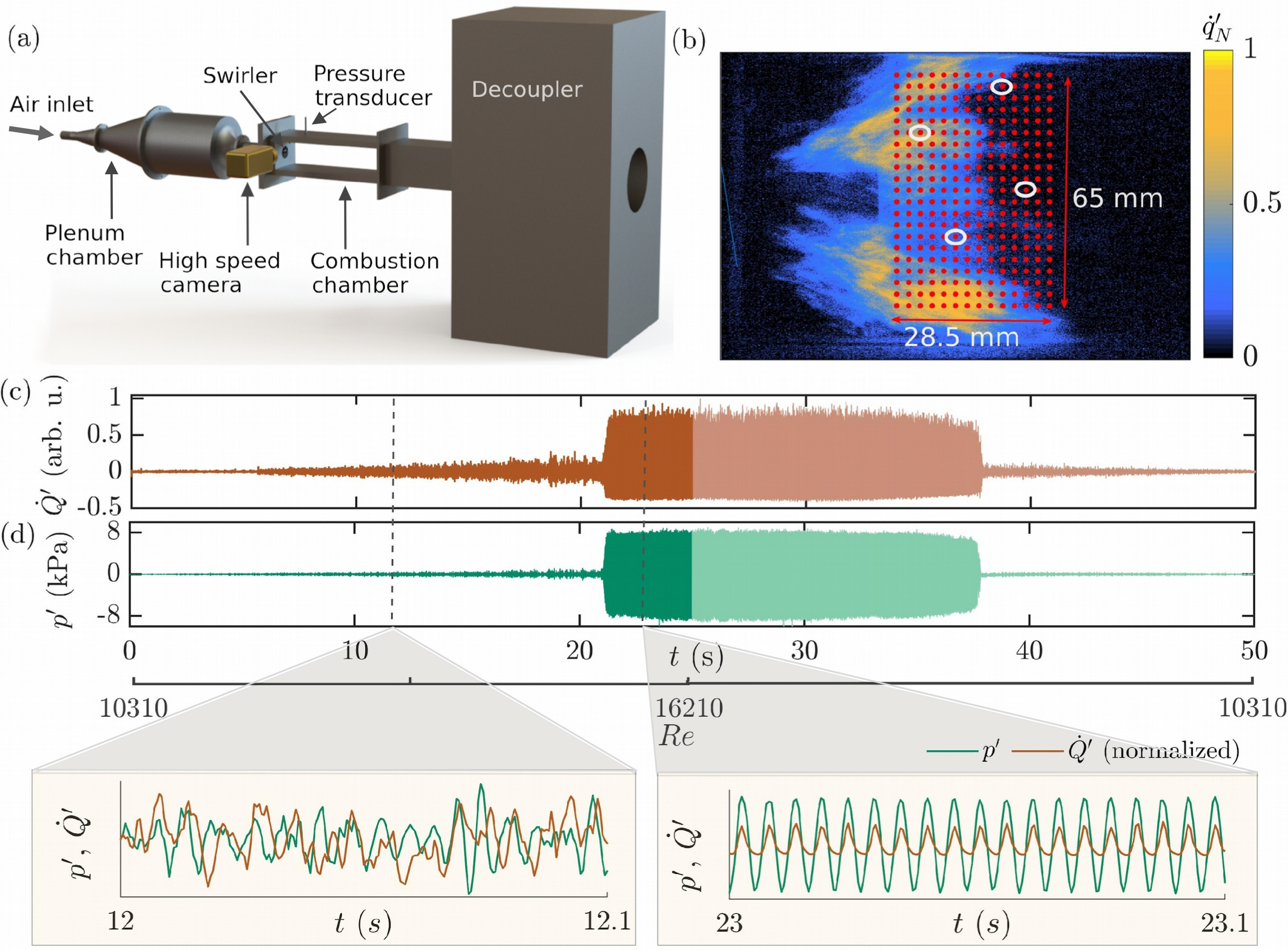}
\caption{\textbf{Schematic of the experimental setup and representative snapshot of the spatiotemporal data and temporal data capturing the abrupt transition.} (\textbf{a}) The turbulent combustor with measurement devices such as a piezoelectric pressure transducer and a camera used for high-speed chemiluminescence imaging. (\textbf{b}) A representative snapshot of the spatiotemporal CH$^*$ chemiluminescence data ($\dot{q}^\prime_N$, normalized with the maximum intensity) during the state of thermo-acoustic instability. Groups of four adjacent pixels are considered oscillators (corresponding to the red dots), and we chose 294 such oscillators from the area of interest. Four oscillators selected arbitrarily, marked with white circles, are representative oscillators used to study the dynamics of individual oscillators.
Time series of (\textbf{c}) global heat release rate fluctuations ($\dot{Q}^\prime$) and (\textbf{d}) acoustic pressure fluctuations ($p^\prime$) obtained from experiments. The data is acquired over a duration of 50 s. In both figures, the darker color indicates the forward path with (increasing in $Re$ from $t$ = 0 s to $t$ = 25 s), and the lighter color indicates the reverse path (decreasing $Re$ from $t$ = 25 s to $t$ = 50 s). Zoomed-in plots for the states before and after the transition to thermo-acoustic instability are also shown.}
\label{fig1}
\end{center}
\end{figure*}

\hspace{1 cm} We perform experiments in a backward-facing step combustor with reactant flow at high Reynolds numbers $(Re > 1\times10^4)$. Figure~\ref{fig1}a shows the schematic of the experimental setup. A detailed description of the setup is provided in Materials and Methods. The fuel-air mixture is spark-ignited in the combustion chamber using an ignition transformer. Experiments are performed by keeping the mass flow rate of fuel ($\dot{m}_f$) fixed at $0.75\pm0.04\ g/s$. The flow rate of air ($\dot{m}_a$) is gradually increased from $7.76\pm0.14\ g/s$ (at $t = 0$ s) up to a maximum value of $12.66\pm0.18\ g/s$ (at $t = 25$ s) at a rate of $0.2\pm0.01\ g/s^2$, resulting in a gradual increase in the value of the Reynolds number. After reaching the maximum value, $\dot{m}_a$ is reduced back to the initial value at the same rate of change. 
Unsteady pressure measurements ($p'$) are acquired using piezoelectric transducers, and high-speed images of flame dynamics are captured using a high-speed CMOS camera by measuring the CH$^*$ chemiluminescence intensity.
% using a Phantom v12.1 high-speed camera simultaneously. 
Here, the chemiluminescence intensity of the CH$^*$ radical represents the heat release rate from the flame \cite{hardalupas2004local, guethe2012chemiluminescence}. 

As the Reynolds number is varied in time, we observe a spontaneous emergence of ordered dynamics (thermo-acoustic instability). We study the spatiotemporal characteristics during the transition to thermo-acoustic instability by analyzing the high-speed CH* chemiluminescence images of the combustion zone. For ease of analysis, we select a grid of size $300\times130$ pixels in the flow field as shown in Fig.~\ref{fig1}b. We consider a group of four adjacent pixels as a local heat release rate oscillator; such a grouping is done as the time series from a single pixel is noisy. We consider 294 such oscillators (arranged as a grid of $21\times14$ oscillators) indicated by the red dots in Fig.~\ref{fig1}b. The heat release rate of a given oscillator ($\dot{q}^\prime$) is computed as the sum of the intensity values of the four pixels that correspond to that particular oscillator. 
The global or integral heat release rate ($\dot{Q}^\prime$)  is then computed as the sum of all the local heat release rate oscillators. As the acoustic pressure is nearly uniform along the reaction zone, the acoustic field is considered as another oscillator, which interacts with all these heat release rate oscillators.

The chemiluminescence images acquired from the turbulent combustor are post-processed along with the unsteady pressure data in order to analyze the spatiotemporal dynamics of the reaction zone. Figures~\ref{fig1}c \& d show the time series of global heat release rate fluctuations and the acoustic pressure fluctuations, respectively, for a duration of 50 s. The Reynolds number of the flow is varied continuously, as indicated in the plot. For low Reynolds numbers, the time series of global heat release rate fluctuations ($\dot{Q}^\prime$) and acoustic pressure ($p^\prime$) exhibit low-amplitude irregular fluctuations (see Fig.~\ref{fig1}, zoomed time series for $t = 12$ to $12.1$ s). These low amplitude aperiodic fluctuations are referred to, in thermo-acoustic parlance, as combustion noise and are actually high-dimensional chaos \cite{tony2015detecting}. This state corresponds to the stable operating regime for the combustor. As the airflow rate is increased further, the combustor exhibits a transition to thermo-acoustic instability with large-amplitude periodic oscillations (Fig.~\ref{fig1}c \& d, zoomed time series for $t = 23$ to $23.1$ s).
% , for a small section of $t = 29$ to $29.2$ s). 
Such a rapid increase in amplitude closely resembles the phenomenon of explosive synchronization. We then reverse the direction of variation of the control parameter at $t = 25$ s by decreasing the airflow rate (see Fig.~\ref{fig1}c \& d, for $t = 25$ to $50$ s). The combustor continues to exhibit thermo-acoustic instability up to $t = 38$ s. As the airflow rate is further decreased, the system abruptly returns to the low amplitude state.

\subsection*{Quantitative characterization of synchrony between individual heat release rate oscillators}
We investigate the interactions across different spatio-temporal scales during the transition from the state of low amplitude aperiodic oscillations (combustion noise) to periodic oscillations (thermo-acoustic instability) by calculating the space-time correlation. Characterization of synchronization between the subsystems is necessary to quantitatively confirm the synchronization transition  at the onset of thermo-acoustic instability.
One of the popular synchrony metrics is the windowed cross-correlation\cite{Golomb:2007,zhang2014explosive,leyva2020inferring}. We calculate the cross-correlation $(C_{ij})$ between pairs of oscillators $(i,j)$, and then calculate the average of $C_{ij}$ of all the pairs of oscillators to obtain $C_{{\dot{q}}^{\prime}{\dot{q}}^\prime}$. 
This procedure is followed for each 1 s long window, and the resultant value of $C_{{\dot{q}}^{\prime}{\dot{q}}^\prime}$ is plotted at the end of each window. Similarly, we estimate the average windowed cross-correlation for the entire duration of the experiment to measure how fast an ordered state is developed from local synchronization among the individual oscillators. The closer the cross-correlation value is to 1, the more synchronized the two signals are.

\begin{figure*}[h!]
\begin{center}
\includegraphics[scale = 0.185] {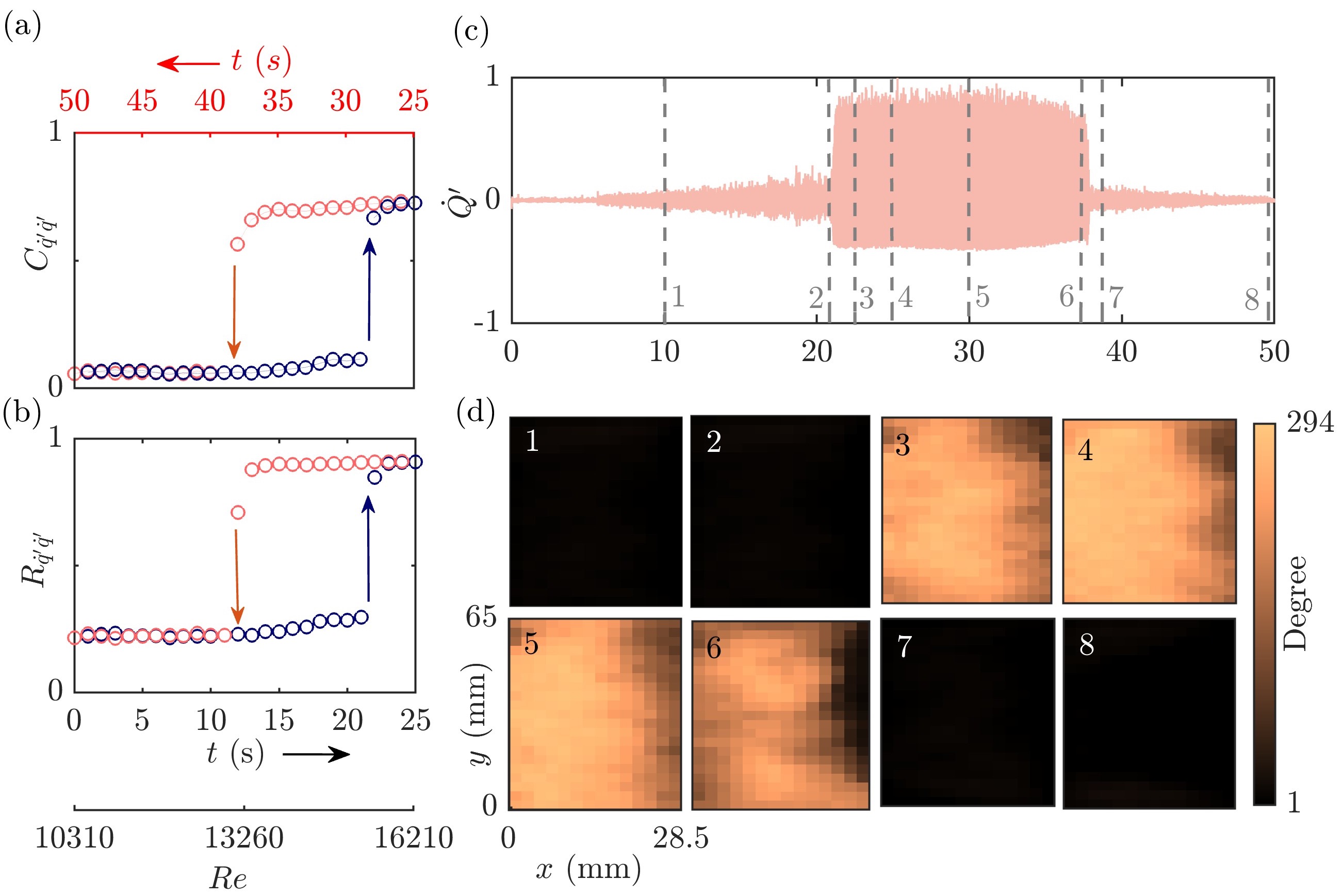}
\caption{\textbf{Quantification of synchronization between local heat release rate oscillators.} (\textbf{a}) Average cross-correlation between pairwise individual heat release rate oscillators ($C_{{\dot{q}}^{\prime}{\dot{q}}^\prime}$). (\textbf{b}) The variation of the Kuramoto order parameter ($R_{{\dot{q}}^{\prime}{\dot{q}}^\prime}$) as a function of time. The blue colored markers indicate the forward path (with time indicated in the bottom abscissa in black color), and the red colored markers represent the reverse path (with time indicated in the top abscissa in red color). In both the plots \textbf{a} \& \textbf{b}, the coherence parameter ($C_{{\dot{q}}^{\prime}{\dot{q}}^\prime}$ or $R_{{\dot{q}}^{\prime}{\dot{q}}^\prime}$) remains low up to time $t$ = 21 s, and then suddenly jumps to a high value ($C_{{\dot{q}}^{\prime}{\dot{q}}^\prime} \sim 0.75\ \&\ R_{{\dot{q}}^{\prime}{\dot{q}}^\prime}\sim 1$) indicating explosive synchronization. A similar abrupt transition is visible at $t$ = 38 s as the value of the coherence parameter abruptly decreases to a value close to zero. The transition displays a hysteresis loop. Note that whenever we consider an average of cross-correlation of pairs of oscillators, we mark the error bar with 95 \% confidence. Based on the cross-correlation between pairs of oscillators, complex networks are constructed, and the evolution of the degree of each oscillator is shown in (\textbf{d})  for selected time instants marked in the timeseries in (\textbf{c}). The time interval for constructing the network is chosen as 1 s. The maximum possible degree is 294. The degree of most of the oscillators is low before the onset of thermo-acoustic instability, and they quickly increase to a higher degree at the onset of oscillatory instability. Then, they also return to low values of degree as in the initial state, as the system returns to the low amplitude chaotic state from thermo-acoustic instability.
 }
\label{fig3}
\end{center}
\end{figure*}

Figure~\ref{fig3}a illustrates that the cross-correlation averaged across all the heat release rate oscillators, $C_{\dot{q}^\prime \dot{q}\prime}$, before the onset of thermo-acoustic instability is close to zero, and at $t$ = 22 s, it suddenly increases to a higher value indicating explosive synchronization. In the reverse path, at $t$ = 38 s, $C_{\dot{q}^\prime \dot{q}^\prime}$ jumps from a higher value to a value close to zero, with the presence of a hysteretic loop. The forward and backward curves do not overlap; instead, they show a hysteretic behavior, pointing to the coexistence of two phases of the system.

% As an alternative method to investigate synchronization, we measure the coherence of the collective motion by using the 
An alternative measure to investigate a synchronization transition is the Kuramoto order parameter\cite{kuramoto1984chemical}, defined as, 
\begin{equation}\label{eq:Kuramoto}
{R(t) = \frac{1}{N} \Bigg| \sum_{j=1}^{N}e^{i{\theta_j}(t)}\Bigg|},
\end{equation}
where, $R(t)$ is the order parameter at time $t$, $\theta_{j}(t)$ is the phase of $j^{th}$ oscillator at time $t$, and $N$ is the total number of oscillators in the network. 
% The Kuramoto order parameter has been used as a good indicator of transition to global synchrony . 
A value of $R(t)$ close to one implies a perfectly synchronized state, while a value close to zero indicates a completely desynchronized state. For the present analysis, we calculate the instantaneous phase of the signal using the Hilbert transform, a widely used method to compute the instantaneous phase in synchronization literature \cite{kuramoto1984chemical,wang2017small}. 

We observe, in Fig.~\ref{fig3}b, that the Kuramoto order parameter computed for the network of heat release rate oscillators, $R_{\dot{q}^\prime \dot{q}^\prime}$, is low before the onset of thermo-acoustic instability, and then it rapidly increases to a value close to unity at $t$ = 22 s, indicating the presence of explosive synchronization. Also, in the reverse path, we observe a similar abrupt transition, but to the state of desynchronized dynamics at $t$ = 38 s. The presence of explosive synchronization is further affirmed by the observation of hysteresis, which is a characteristic feature of first-order synchronization transitions.

To further demonstrate the spatial dynamics during the first-order synchronization transition for the spatially distributed oscillators, we construct complex networks using the cross-correlation between pairs of oscillators. 
% A degree position matrix (from the adjacency matrix) can be constructed to represent. 
If a pair of oscillators has a $C_{\dot{q}^\prime \dot{q}^\prime}(t)$ greater than a threshold value of $0.6$, we connect them. Note that similar results can be obtained for a range of correlation thresholds (discussed in the Appendix). Using such a criterion for connectivity in the network, the degree of each oscillator is calculated, where the degree of an oscillator is the total number of connections of that particular oscillator. The spatial distributions of the degree calculated for a network constructed from the timeseries of short duration (estimated for a time window of 1 s) are shown for selected conditions in Fig.~\ref{fig3}c. The evolution of the degree of the cross-correlation network in Fig.~\ref{fig3}d also shows an abrupt change in behavior around $t$ = 22 s and $t$ = 38 s.  
All the above quantitative results prove that an explosive synchronization happens between the heat release rate oscillators, with the emergence of a hysteresis loop. 

\subsection*{Quantitative characterization of synchrony between the global heat release rate and the acoustic pressure oscillations}
Next, we investigate the interactions between the acoustic pressure and the heat release rate during the transition to thermo-acoustic instability. The time series of the global heat release rate (GHRR) or the integral heat release rate is calculated by adding the intensities of local heat release rates (LHRR) (here, 294 local oscillators). We used $\dot{q}^\prime$ to denote fluctuating heat release rate from the local oscillators; here, we use $\dot{Q}^\prime$ for global heat release rate fluctuations.
%Since we are dealing with a spatially extended system with a mean flow, there will be a time lag that needs to be accounted for in our analysis. As a result, the time lag in synchronization of each oscillator with respect to the pressure must be considered. In order to account for this, we are using the following methods to analyze the characteristics of synchrony between the global heat release rate and the acoustic pressure. 
                   
\begin{figure*}[h!]
\begin{center}
\includegraphics[scale = 0.175] {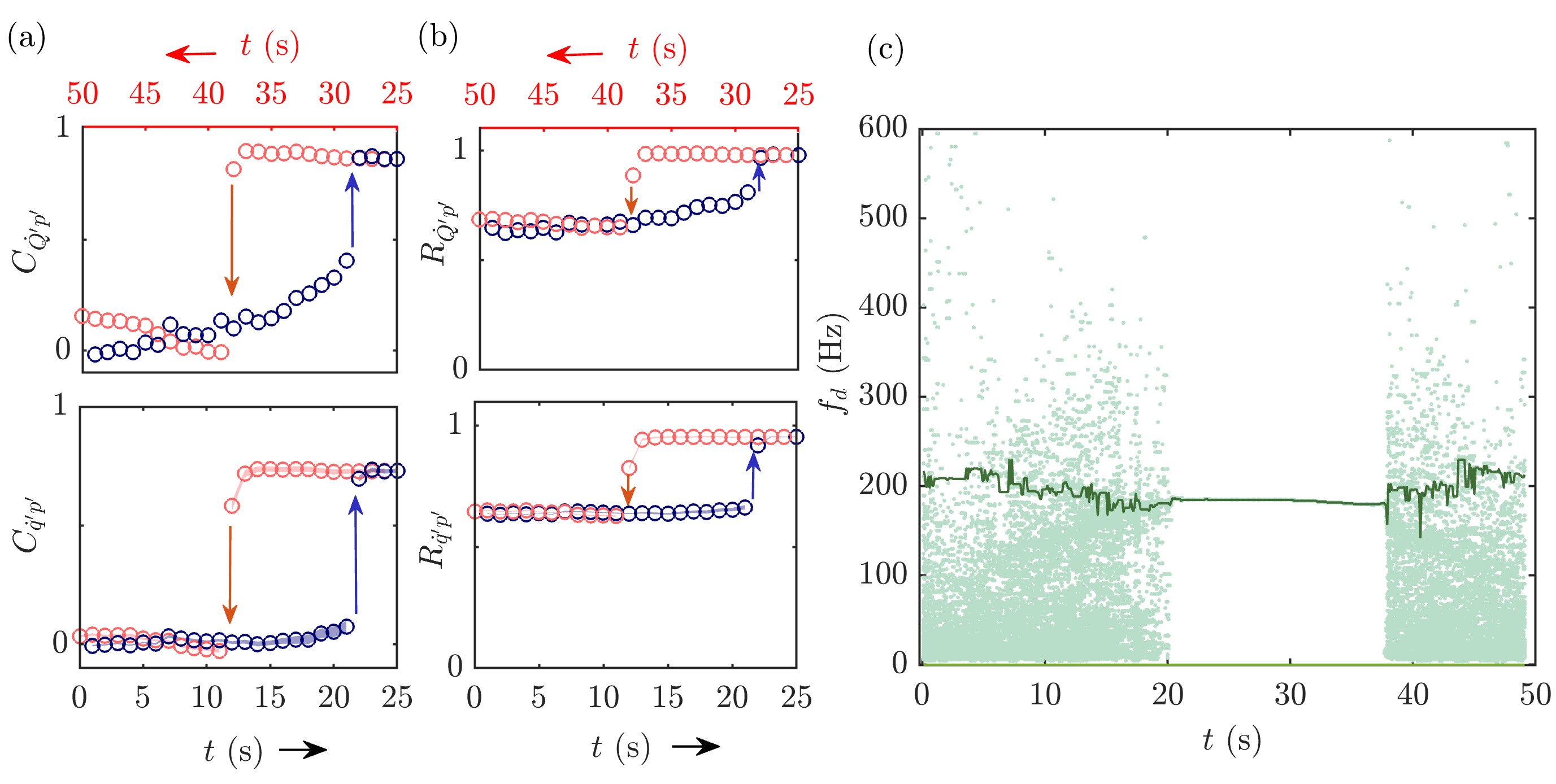}
\caption{\textbf{Synchronization between the acoustic pressure and the heat release rate fluctuations.} (\textbf{a}) The variation of cross-correlation between the global heat release rate and the acoustic pressure fluctuations. The value of cross-correlation gradually increases till $t$ = 21 s  and then jumps to a higher value. In the backward path, at $t$ = 38 s, it jumps from a higher value of the synchronization parameter to a lower value. The average cross-correlation between the local heat release rate oscillators and the acoustic pressure fluctuations is shown at the bottom. The value of correlation prior to transition is relatively high for the case of integral heat release rate compared to the case of local heat release rate oscillators. (\textbf{b}) The variation of Kuramoto order parameter between the global heat release rate and the acoustic pressure fluctuations. (\textbf{c}) The evolution of dominant frequencies in the power spectrum for different $Re$. The thick dark green colored line represents the variation in the dominant frequency of acoustic pressure, and the other light green colored dots represent the dominant frequency of each of the 294 oscillators. We observe that before $t$ = 22 s, both the pressure and GHRR fluctuations have distinct and noisy frequencies. At $t$ = 22 s, a sudden and abrupt locking of these frequencies to a single value up to $t = $ 38 s occurs, suggesting mutual explosive synchronization between these signals in the thermo-acoustic system.}
\label{fig4}
\end{center}
\end{figure*}

We use the windowed cross-correlation method and calculate the global cross-correlation between the acoustic pressure and global heat release rate $(C_{\dot{Q}^\prime p^\prime})$,  within a window of 1 s (Fig.~\ref{fig4}a top figure). 
% We then compute the average of $R_{p\dot{q_j}}$ to determine the synchronization between acoustic pressure and global heat release rate for each time window $t$. 
We also calculate the pairwise cross-correlation between each local heat release rate oscillator and the acoustic pressure $(C_{p^\prime \dot{q}^\prime_j}),\ j = 1,2...\ N$.
The average of $(C_{p^\prime \dot{q} ^\prime_j})$ over all oscillators ($C_{p^\prime\dot{q}^\prime}$) is then computed (shown in Fig.~\ref{fig4}a bottom figure). Then the same procedure is repeated for each 1 s long window.
In the case of synchronization between the global variables, $C_{\dot{Q}^\prime p^\prime}$ (Fig.~\ref{fig4}a), the integral heat release rate and the acoustic pressure fluctuations exhibit a gradual synchronization up to a high value of cross-correlation and then suddenly synchronize to the fully synchronized state just at the onset of thermo-acoustic instability. In contrast, the average cross-correlation between individual oscillators and acoustic pressure jumps abruptly to a higher value at $t$ = 22 s, indicating explosive synchronization. During the reversal of the parameter, the system exhibits an abrupt transition to a desynchronized state at time $t$ = 38 s. This confirms that explosive synchronization happens between the pressure and the local heat release rate fluctuations along with the emergence of a hysteresis loop, while the global heat release rate starts to synchronize gradually with acoustic pressure before the abrupt jump.

We then investigate the variation of Kuramoto order parameter during this transition. Due to the aperiodic nature of the time series during the state of combustion noise, defining the phase of each oscillator is complicated. In addition, the acoustic pressure and heat release rate fluctuations are different chaotic oscillators. This makes it challenging to estimate the Kuramoto order parameter in a consistent and meaningful manner, which is evident in the Kuramoto coherence measured between the pressure and heat release rate fluctuations in Fig.~\ref{fig4}b. As we can observe from Fig.~\ref{fig4}b, the Kuramoto order parameter does not give a clear picture of synchronization between the acoustic pressure and the heat release rate fluctuations.
% ; it is not the best method for estimating the synchronization between two different types of signals.
Hence, we will refrain from using the Kuramoto order parameter to draw any inferences in this study.
While Kuramoto coherence is a promising measure for quantifying synchronization in a system of identical oscillators, it may not be the most appropriate measure in situations such as the one here involving nonidentical aperiodic oscillators \cite{pikovsky2003synchronization}. Alternative measures, such as mutual information or cross-correlation, may be more suitable for measuring coherence between different systems. These measures have been successfully used to quantify synchronization between non-identical systems of oscillators. Therefore, the choice of synchronization measure should depend on the specific characteristics of the system being studied.

The coupled interaction of the integral heat release rate and the acoustic pressure signals can further be understood through a frequency domain analysis. Figure~\ref{fig4}c shows a plot of the evolution of the dominant frequencies of all the oscillators, demonstrating the dominance of one single frequency as the system dynamics evolves in time towards a synchronized state. We notice the presence of a dominant frequency around 220 Hz in the acoustic pressure signal and a distribution of frequencies for LHRR, during the state of combustion noise. During the transition of the system dynamics from combustion noise to limit cycle oscillations, we observe a sudden and abrupt transition to a common and very sharp peak frequency of 190 Hz (see Appendix for the evolution of the amplitude spectrum in time). This type of abrupt frequency switching from broadband to a narrow peak is a feature of explosive synchronization transition.

\begin{figure*}[h!]
\begin{center}
\includegraphics[scale = 0.138] {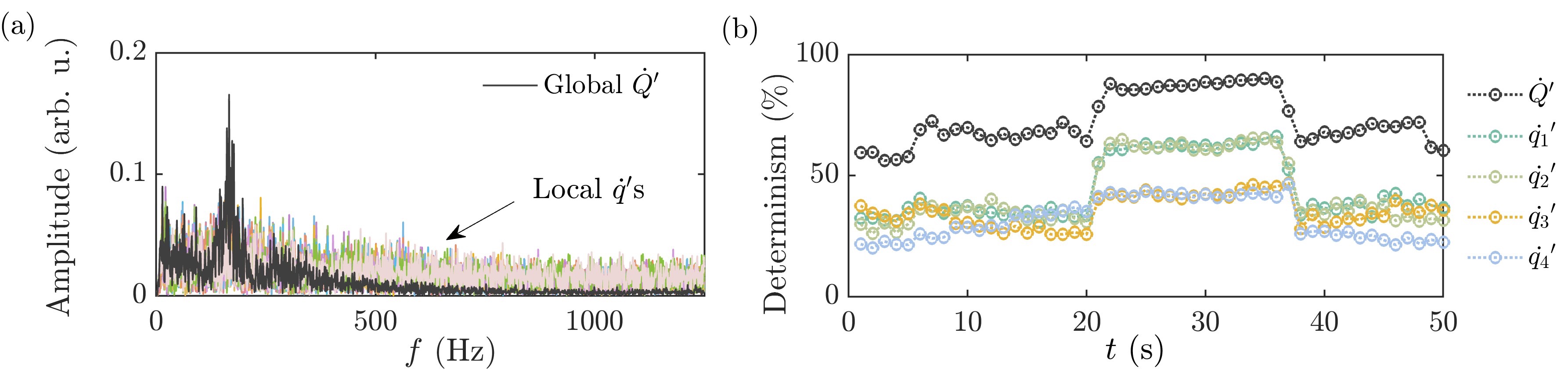}
\caption{\textbf{A comparison of the values of determinism between GHRR signal and LHRR signals.} (\textbf{a}) Amplitude spectra of a few local heat release rate oscillators selected arbitrarily and that of global heat release rate oscillation calculated through the fast Fourier transform. (\textbf{b}) The percentage of determinism of GHRR (black circles) and four other LHRR signals. We find that GHRR shows a very high value of determinism compared to all the individual oscillators.}
\label{fig5}
\end{center}
\end{figure*}

Examining Fig.~\ref{fig4}, it is evident that the interaction between the integral heat release rate and the acoustic pressure fluctuations (Fig.~\ref{fig4}a, top figure) is different from the interaction between the local heat release rate and the acoustic pressure fluctuations (Fig.~\ref{fig4}a, bottom figure). We analyze this further by comparing the amplitude spectra of the integral heat release rate and a number of randomly selected local heat release rate oscillators just before the explosive transition, i.e., in the window starting at 21\textsuperscript{st} second. Using the fast Fourier transform, the amplitude spectra of the integral heat release rate and a few randomly chosen local heat release rates are plotted (Fig.~\ref{fig5}a).

In Fig.~\ref{fig5}a, the black curve indicates the amplitude spectrum for the integral heat release rate, while the other colored lines are of a few randomly selected local heat release rate oscillators. We can clearly observe that the integral heat release rate has a sharper peak with significantly less contribution from higher frequencies when compared to individual heat release rate oscillators. The relatively high deterministic nature of integral heat release rate oscillations is further highlighted in Fig.~\ref{fig5}b using a measure known as determinism calculated from quantifying the recurrences of trajectories in phase space \cite{Eckmann_1987,webber2015recurrence}. The details of the calculation of determinism are given in Appendix. Figure~\ref{fig5}b compares the percentage of determinism of the GHRR signal and four LHRR signals, chosen from the locations marked with white circles in Fig. \ref{fig1}b. The GHRR signal shows higher determinism, both during the states of combustion noise and thermo-acoustic instability. This corroborates the higher level of synchrony between the GHRR and acoustic pressure signals in comparison to the LHRR and the acoustic pressure signals, as illustrated in Fig.~\ref{fig4}a. 
% Here, the determinism is calculated by estimating the recurrence of the heat release rate signals \cite{webber2015recurrence}.

In summary, the acoustic field synchronizes with the global heat release rate oscillator gradually and then abruptly attains complete synchronization. Initially, when the value of the control parameter is low, both the correlation between individual heat release rate oscillators and with the acoustic pressure fluctuations are weak. As we increase $Re$, the global heat release rate and the acoustic field start to synchronize gradually with a consistent increase in the cross-correlation. They synchronize gradually and reach a critical value of $C_{\dot{Q}^\prime p^\prime}$, triggering all the individual heat release oscillators to synchronize with each other and with the acoustic field. Every oscillator getting synchronized with each other manifests as a spontaneous onset of a fully synchronized state, and the entire network activates suddenly, like turning on a switch. 

\subsection*{Model}
We model the temporal and spatiotemporal synchronization leading to ordered dynamics in a complex system where multiple subsystems interact to cause spontaneous synchronization. We construct a network of $N$ Rössler oscillators (Eq. \ref{eq0}) arranged as a square grid. Here, we consider the spatially distributed Rössler oscillators to be analogous to the spatiotemporal heat release rate oscillations in the turbulent reactive flow system and couple them diffusively. A chaotic Van der Pol oscillator, VDP, (Eq. \ref{eq2}) is used analogous to the acoustic field and hence we couple it to all the Rössler oscillators. We adopt the model used by Godavarthi et al. \cite{godavarthi2020synchronization} and modify it with an adaptive coupling scheme \cite{zhang2016model} to capture explosive synchronization.
\begin{eqnarray}\label{eq0}
 \centering
      &\dot{x}_i  = -w_{i}y_i-z_i,\nonumber\\  
      &\dot{y}_i  = w_{i}x_i+ay_i+R^\beta\bigg(\mu_{vy}(v-y_i)+\mu_{yy}\sum\limits_{j\in \delta (y_i)}^{}(y_j-y_i)\bigg),\nonumber\\  
      &\dot{z}_i = b+z_{i}(x_i-c),
 \end{eqnarray}
% \begin{equation*}
%     \dot{x}_i & = -w_{i}y_i-z_i,
% \end{equation*}
% \begin{equation}\label{eq1}
%     \dot{y}_i & = w_{i}x_i+ay_i+R^\beta\bigg(\mu_{vy}(v-y_i)+\mu_{yy}\sum_{j\epsilon\delta (y_i)}^{}(y_j-y_i)\bigg),
% \end{equation}
% \begin{equation*}
%     \dot{z}_i &= b+z_{i}(x_i-c),
% \end{equation*}
 \begin{eqnarray}\label{eq2}
 \centering
 &\dot{u} = v, \nonumber\\  
 &\dot{v} = 0.1(1-u^2)v-u^3+\cos t+(1+R^\beta)\bigg(\mu_{yv}\sum\limits_{k=1}^{N}(y_k-v)\bigg),
 \end{eqnarray}
% \begin{equation}\label{eq2}
%     \dot{u} = v,
% \end{equation}
% \begin{equation*}
%     \dot{v} &= 0.1(1-u^2)v-u^3+\cos t+R^\beta\bigg(\mu_{yv}\sum_{k=1}^{N}(y_k-v)\bigg),
% \end{equation*}
where the individual Rössler oscillators are defined by the parameters $a,b$ and $c$. The system exhibits chaotic behavior when $a$ is set to $0.165$, $b$ is set to $0.2,$ and $c$ to $10$. The total number of Rössler oscillators is denoted as $N$. Intra-Rössler coupling operates through the second coordinate of the oscillators ($y_i$), forming connections to their nearby counterparts within a distance of $\sqrt{2}$ units. This coupling strength is represented as $\mu_{yy}$. The neighborhood around the $i^{th}$ Rössler oscillator is indicated as $\delta (y_i)$ in Eq. \ref{eq0}. We model the complexities arising due to the underlying turbulent flow as the variability of natural frequencies of these oscillators. Consequently, non-identical Rössler oscillators are adopted by assigning natural frequencies (\(\omega_i\)) from a Gaussian distribution characterized by a mean of 0.8 and a standard deviation of 0.05. 
% Equation (\ref{eq2}) corresponds to a chaotic VDP oscillator. 

The system of Rössler oscillators and the chaotic VDP oscillator are coupled bidirectionally (between variable $y$ and $v$) with strengths of $\mu_{vy}$ and $\mu _{yv}$, corresponding to the influence of the VDP oscillator on a Rössler oscillator and vice versa. Following Zhang et al. \cite{zhang2016model}, we use the factor $R^{\beta}$ in the coupling terms in Eq. \ref{eq0} and Eq. \ref{eq2} to adaptively control the coupling strength. Here, $R$ is the order parameter of the Rössler system as explained in Eq. \ref{eq:Kuramoto}. 
% given by 
% \begin{equation}
% {R(t) = N^{- 1} \Bigg| \sum_{j=1}^{N}e^{i{\theta_j}(t)}\Bigg|}
% \end{equation}
When $\beta$ = 0, the model exhibits the standard continuous transition \cite{godavarthi2020synchronization}, whereas, $\beta$ = 2 results in a discontinuous transition. The transition changes from continuous synchronization to explosive synchronization with the change in $\beta$. Here, $\beta$ acts like a parameter controlling the nonlinear variation of the coupling strengths.

%We hypothesize that 
The coupling strength $\mu _{yy}$ quantifies the interaction between the Rössler oscillators similar to the interaction between the local fluctuations in the heat release rate, which occurs through turbulent diffusion during combustion \cite{godavarthi2020synchronization}. The chaotic VDP oscillator is analogous to the global acoustic field and is therefore coupled to each Rössler oscillator. 
We further set the relations among the coupling strengths $\mu _{yy}$, $\mu _{yv}$, and $\mu _{vy}$ based on experimental observations. 
%Moreover, Godavarthi et al. \cite{godavarthi2018coupled} showed that the unsteady flame dynamics exerts a stronger influence on the acoustic field than vice versa. Hence, we vary $\mu _{yv}$ and $\mu _{vy}$ as per the relation $\mu _{yv}$ = 1.2$\mu _{vy}$. 
We vary $\mu _{yv}$ and $\mu _{vy}$ as per the relation $\mu _{yv}$ = 0.004$\times$ (0.8 + $\mu _{yy}$) and $\mu _{vy}$ = 1.5$\times$(0.8 +$\mu _{yy}$). When the heat release rate oscillations synchronize among themselves and with the acoustic pressure fluctuations, coherent structures or patterns emerge in the flow field. These structures, in turn, enhance the interaction among the spatially distributed local heat release rate oscillations, and we represent these interactions using $\mu_{yy}$.

%We fix $N = 100$ for the simulations and both the coupling strengths $\mu _{yv}$ and $\mu _{vy}$ are increased linearly. $\mu _{yy}$ is gradually increased from -0.8 in steps of $10^{-1}$ along with $\mu _{yv}$ from 0 in steps of $10^{-3}$. 
We fix $N = 100$ for the simulations, and the coupling strength $\mu _{yy}$ is gradually increased from -0.8 in steps of $10^{-1}$.
In order to check for a first-order synchronization transition, simulations are performed by increasing and then decreasing the coupling strengths. %Data from each step is acquired after discarding a long transient for further processing.
\begin{figure*}[h!]
\begin{center}
\includegraphics[scale = 0.19] {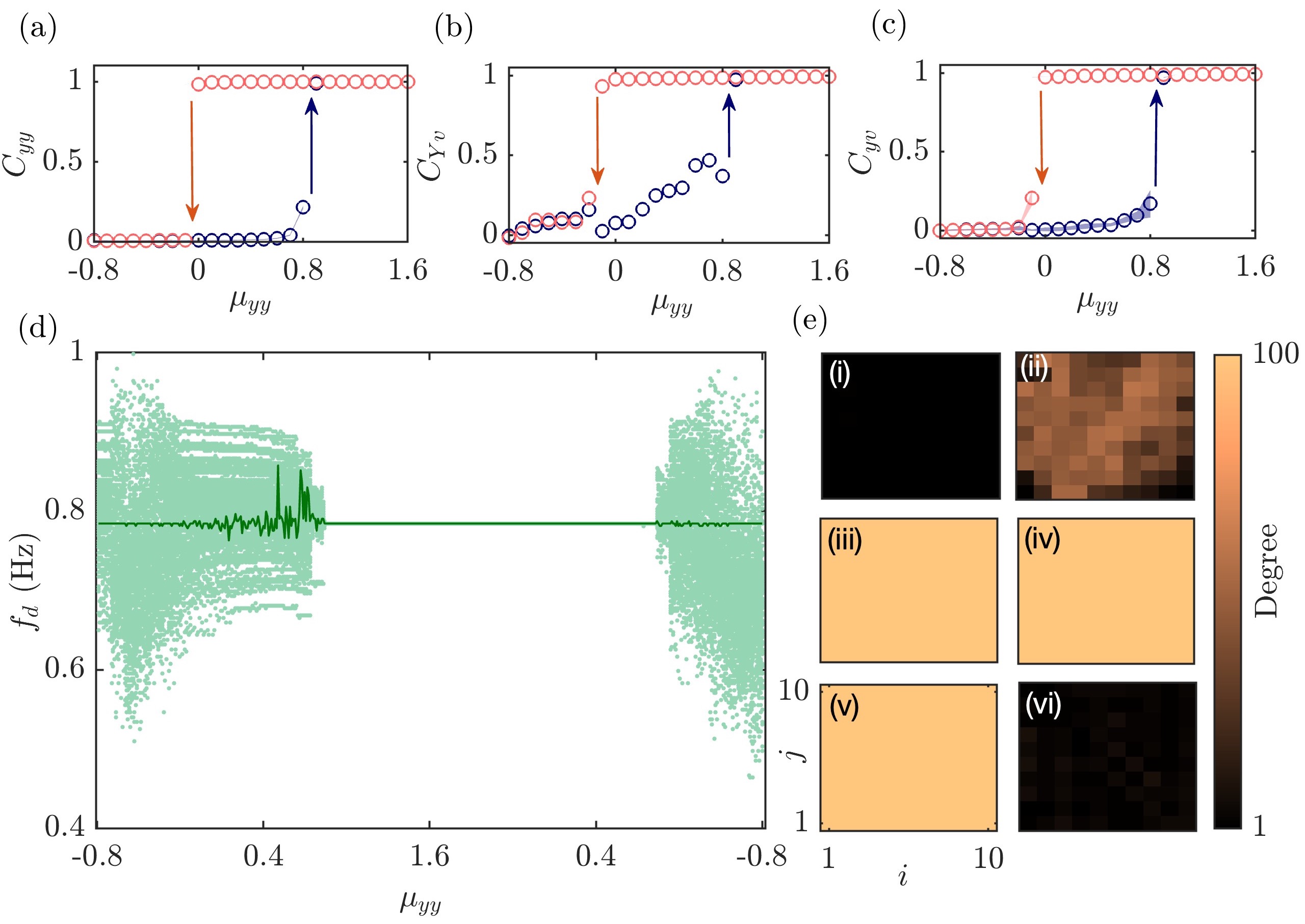}
\caption{\textbf{Synchronization quantification of  the model.}  (\textbf{a}) The average cross-correlation between all the pairs of Rössler oscillators as a function of intra-Rössler coupling strength $\mu_{yy}$. (\textbf{b}) The average cross-correlation between the mean-field of Rössler oscillators and VDP oscillator as a function of $\mu_{yv}$. (\textbf{c}) The average cross-correlation between each Rössler oscillator and VDP oscillator as a function $\mu_{yy}$. The shaded region represents the uncertainty obtained by calculating the average over many realizations. For this type of coupling in the network of oscillators, the synchronization among the individual oscillators and that between the mean field and the VDP exhibit a very similar nature of synchronization as observed in our experiments. (\textbf{d}) The evolution of dominant frequencies in the power spectrum for different $\mu_{yy}$. The thick green line represents the variation of the dominant frequency of the VDP oscillator, and the other colored dots represent the dominant frequency of each of the 100 Rössler oscillators. (\textbf{e}) The evolution of the degree of the Rössler oscillators is shown here at the value of $\mu_{yy}$ at $(i)$ -0.8, $(ii)$ 0.8, $(iii)$ 0.9, $(iv)$ 1.6, $(v)$ 0, and $(vi)$ -0.1. The maximum possible degree is 100. The degree of most of the oscillators is low before the onset of synchronization, and they quickly increase to a higher degree at the onset of synchronization. Then, they return to low values of degree as in the initial state, as the system returns to the low amplitude chaotic state from the synchronized state.}
\label{fig6}
\end{center}
\end{figure*}
We determine the level of synchronization between pairs of individual Rössler oscillators (Fig.~\ref{fig6}a) and between each Rössler oscillator and VDP oscillator (Fig.~\ref{fig6}c), using the average of cross-correlation as the order parameter. Following Godavarthi et al. \cite{godavarthi2020synchronization}, the sum of the variable $y$ of Rössler oscillators is considered analogous to the global heat release rate with which we calculate the level of synchronization with the VDP oscillator (Fig.~\ref{fig6}b) by calculating the cross-correlation between them. 

In Fig.~\ref{fig6}a, the order parameter ($C_{yy}$) stays low as we increase the coupling parameter up to a coupling parameter $\mu _{yy}$ = 0.8, after which it abruptly jumps to a higher value, indicating explosive synchronization. While decreasing the coupling strength, we observe a hysteresis region, a feature of explosive synchronization. This plot is similar to Fig.~\ref{fig3}a from experimental data. 
We also estimate the correlation between the Rössler oscillators with the VDP oscillator. As we increase $\mu _{yy}$, the coupling with VDP oscillator ($\mu _{yv}$ and $\mu _{vy}$) also increases accordingly. In Fig.~\ref{fig6}b the order parameter increases gradually, and it suddenly jumps to a higher value around $\mu_{yy}$ = $0.8$. Further, as we decrease the coupling strength, a hysteresis region is observed. The average cross-correlation between the VDP and individual Rössler oscillators resembles the behaviour observed in the experiments (Fig. \ref{fig6}c).
The dynamics illustrated in these plots (Fig. \ref{fig6}b \& c) are similar to Fig.~\ref{fig4}a. Hence, we can say that the interaction of individual Rössler oscillators among themselves and with the VDP oscillator is similar to that between individual heat release rate oscillators among themselves and with the acoustic pressure fluctuations that we obtained in our experiment. 

Here, the Van der Pol oscillator, analogous to the acoustic field, acts as the global communicator. Initially, when the control parameter, $\mu _{yy}$, is very small, both intra-Rössler and Rössler-VDP couplings are weak. As we increase $\mu _{yy}$ and in turn, $\mu_{yv}$, analogous to the coupling between global heat release rate and acoustic field, the connection between the mean field of the Rössler system and the VDP becomes stronger. They synchronize gradually and reach a critical value, triggering all the individual Rössler oscillators to synchronize with each other and with the VDP. The global variables appear to show a gradual synchronization followed by a spontaneous synchronization of the entire network of oscillators. 

The phenomenon of frequency locking is also observed in the model (Fig.~\ref{fig6}d). The individual Rössler and VDP oscillators have different frequencies, which lock to a single frequency at the onset of synchronization. The oscillators are locked in this frequency until they become desynchronized again. This transition is sudden and shows explosive behaviour, just like the experimental results. The evolution of the spatial distribution of the degree of oscillators also shows a sudden transition just before and after synchronization, transitioning from a low value of degree to a very high value (Fig.~\ref{fig6}e) and back.

\section*{Discussions}
We provide experimental evidence of explosive synchronization in a real-world complex system, i.e., a turbulent reactive flow system. We detect a sudden spatiotemporal transition from a desynchronized chaotic state to a globally synchronized state, where all the local heat release rate oscillators synchronize and oscillate periodically. While the synchronization among these local heat release oscillators is abrupt, the global heat release rate starts to synchronize gradually with the acoustic field (another subsystem) and then jumps to a complete synchronization suddenly after reaching a certain level of synchronization. 
% Also, the synchronization between acoustic pressure and heat release rates gradually increases and then abruptly jumps to a higher value of the order parameter. We used a model consisting of a network of $N$ Rössler oscillators and a single Van Der Pol oscillator and arrived at similar results obtained from the experiment. Using the control parameter $\beta$, we were able to trigger a first-order synchronization transition in the model, which was previously continuous. 
Such an abrupt transition to a state of complete synchronization among the subsystems of a complex system would be dangerous in real-world engineering systems as well as natural systems. In engineering systems such as the thermo-fluid system discussed in this study, explosive synchronization to ruinously high amplitude self-sustained oscillations \cite{bhavi2023abrupt,roy2021flame,pavithran2022tipping} can damage the machinery and jeopardize its normal operations resulting in a huge loss of revenue \cite{lieuwen2005combustion}. Explosive synchronization in biological systems, such as the onset of epileptic seizures, is detrimental \cite{adhikari2013localizing,wang2017small}. 
Explosive synchronization appears as a sudden huge change in the state of the system, and early detection of such abrupt transitions is often challenging because the system may not show any significant change in dynamics before the jump. Hence, understanding the dynamics of such transitions is essential for preventing them, mitigating their effects, and developing early warning systems. To the best of the authors' knowledge, this is the first experimental study exploring the spatio-temporal dynamics during the explosive synchronization in a real-world complex system.

\section*{Materials and methods} 
% \begin{bibunit}[naturemag]
\subsection*{Details of the experiments}
We performed experiments on a backward-facing step combustor with reactant flow at high Reynolds numbers. Figure~\ref{fig1}a shows the schematic of the experimental setup. It consists of a plenum chamber, a burner tube of 40  mm diameter, a fixed vane swirler for flame-holding, and a combustion chamber of cross section 90 mm $\times$ 90 mm and length 800 mm. The swirler has a length of 30 mm and has eight fixed vanes bent at an angle ($40^{\circ}$ about the axis) to impart radial momentum to the incoming axial flow. Swirler has a centre body of length 30 mm and diameter of 16 mm at the outer end of the swirler. One end of the swirler is aligned with the exit plane of the burner, and it is located such that the centre body is positioned 30 mm inside the combustion chamber. The central shaft is used to deliver fuel into the combustion chamber through four radial injection holes having a diameter of 1.7 mm located 100 mm upstream from the swirler.
To prevent flashback, a circular disk (thickness = 2 mm, diameter = 40 mm) with 300 holes 
% of diameter 1.7 m 
is inserted 30 mm downstream of the fuel-injection location. 
A blowdown mechanism is used to supply air from high-pressure tanks. The compressed air passes through a moisture separator before entering the plenum chamber. 

Unsteady pressure measurements ($p'$) are acquired using piezoelectric transducers (model number.: PCB103B02) with a sensitivity of 217.5 mV/Pa. The transducers are flush-mounted through the pressure ports with Teflon adapters and with semi-infinite wave guides of length 6 m. A quartz window is provided in the combustor (400 mm $\times$ 90 mm) allowing optical access to the combustion chamber. High-speed images of flame dynamics (applying a narrowband CH$^*$ filter with a peak at 432 nm, 10 nm FWHM) are captured with a Phantom v12.1 high-speed camera, simultaneously with the pressure measurements. The chemiluminescence intensity represents the heat release rate from the flame \cite{hardalupas2004local, guethe2012chemiluminescence}. We acquire acoustic pressure and chemiluminescence data for 50 s at a sampling frequency of 4 kHz and 2 kHz, respectively. A 16-bit analog-to-digital conversion card (NI-6143) is used for data acquisition. 
% The configuration helps to prevent the transducers from excess heating and also ensured that the phase correction required is less than $2^\circ$.  

The supply of fuel and air into the combustion chamber is controlled using mass flow controllers (Alicat Scientific, MCR Series) with a measurement uncertainty of $\pm(0.8 \%$  of reading $+ 0.2\%$  of full scale). Liquefied petroleum gas (LPG) is used as the fuel, which is $60\%$  $C_{4}H_{10}$ and $40\%$  $C_{3}H_{8}$ by volume. We ignite at low flow rates of air and fuel and then increase the flow rates up to $380$ SLPM (air flow rate) \& $24$ SLPM (fuel flow rate) in an interval of $5$ s. Subsequently, the fuel flow rate ($\dot{m}_f$) was held fixed at $24$ SLPM (0.75 g/s) and the airflow rate ($\dot{m}_a$) was gradually increased from $380$ SLPM (7.76 g/s) up to a maximum value of $620$ SLPM (12.66 g/s) in 25 s at a rate of $9.6$ SLPM/s (0.2 g/s$^2$), leading to progressive increase in Reynolds number. After attaining the maximum, the air flow rate is decreased back to $380$ SLPM at the same rate from $t = 25$ s to $t = 50$ s. The Reynolds number is computed using the expression $Re = 4\dot{m}/(\pi \mu D)$, where $\dot{m}= (\dot{m}_a + \dot{m}_f)$ is the mass flow rate of the fuel-air mixture, $D$ is the diameter of the swirler and $\mu$ is the dynamic viscosity of the fuel-air mixture at the operating conditions. The Reynolds number is modified to account for the changes due to the changes in viscosity for the varying air–fuel ratios; the procedure can be found in Wilke \cite{wilke1950viscosity}.

\subsection*{Calculation of cross correlation}
\hspace{1 cm}There are multiple ways to quantify synchrony between two signals. Cross-correlation is one of the most common synchrony measures \cite{Golomb:2007}. Cross-correlation measures the similarity between a vector \textbf{x} and shifted (lagged) copies of a vector \textbf{y} as a function of the lag. 
% It can identify directionality between two signals such as a leader-follower relationship in which the leader initiates a response which is followed by a follower.  
Cross-correlation is a standard method of estimating the degree to which two time series are correlated. 
% The true cross-correlation sequence of two jointly stationary random processes, $x_n$ and $y_n$, is given by
% \begin{eqnarray}\label{eq10}
%  \centering
%     R_{xy} (m) = E\{x_{n+m}y^{*}_{n}\} = E\{x_{n}y^{*}_{n-m}\}
%  \end{eqnarray}

%  where $-\infty < n <\infty $, the asterisk denotes complex conjugation, and $E$ is the expected value operator.
 
% \begin{eqnarray}\label{eq10}
%  \centering
%     \hat{R}_{xy} (m) = 
%     \left
%     \{
%     \begin{array}{cc}
%     \sum\limits_{n=0}^{N-m-1}x_{n+m}y^*_n, &  m \geq k\\
%     \\
%     \hat {R}^*_{xy}(-m), &  m < k\\
%     \end{array}
%     \right.
%  \end{eqnarray}

% The output vector, c, has elements given by
% \begin{eqnarray}\label{eq11}
%  \centering
%     c(m) = \hat{R}_{xy} (m-N) = E, & m = 1,2,3....2N-1.
%  \end{eqnarray}
% By using the cross correlation function, the time lag with maximum correlation is calculated the signal pairs and the second signal is phase shifted with this lag. Now, the correlation coefficient is calculated for the signal $x_n$ and new phase shifted signal $y_n$. 
The correlation coefficient of two random variables is a measure of their linear dependence \cite{Benesty2009}. We consider the cross-correlation of two time series data without any lag (Pearson correlation). If each variable has $N$ scalar observations, then the Pearson correlation coefficient is defined as,
\begin{eqnarray}\label{eq12}
 \centering
    \rho(\textbf{x},\textbf{y}) = \frac{1}{N-1}
    \sum\limits_{i=1}^{N} \left(\frac{\textbf{x}_i-\mu_\textbf{x}}{\sigma_\textbf{x}}\right) \left(\frac{\textbf{y}_i-\mu_\textbf{y}}{\sigma_\textbf{y}}\right)
 \end{eqnarray}
where $\mu_\textbf{x}$ and $\sigma_\textbf{x}$ are the mean and standard deviation of $\textbf{x}$, respectively. and analogously for $\textbf{y}$.
 
 \subsection*{Estimation of determinism through recurrence analysis}
Recurrence analysis is used to examine recurrent behavior or patterns in a time series. It involves identifying similar or repeated patterns within the data, which can provide insights into the underlying dynamics, stability, or predictability of the system dynamics \cite{webber2015recurrence}.
The first step is to construct a recurrence plot to visualize the recurrence patterns in the data \cite{Eckmann_1987}. The recurrence matrix is a binary matrix that indicates whether pairs of data points are considered similar or recurrent based on a defined threshold or similarity criterion. It reveals the recurrent structures or patterns as diagonal lines or clusters \cite{Eckmann_1987}. 

To construct a recurrence plot, first, we preprocess the time series data to detrend or normalize the data. This step ensures that the recurrence plot accurately reflects the underlying dynamics. Then, apply time delay embedding to reconstruct the phase space of the system, which involves creating higher-dimensional vectors by taking multiple lagged copies of the time series. The embedding parameters, such as embedding dimension and time delay, are typically determined using methods such as average mutual information or autocorrelation function and false nearest neighbors algorithm. Then, we compute the Euclidean distance between pairs of embedded vectors \cite{cao1997practical}. 
% The choice of the distance measure depends on the characteristics of the data and the specific objectives of the analysis. 
% We need a recurrence threshold to determine when two vectors are considered recurrent. The threshold can be based on a fixed distance value or a percentage of the maximum distance or by fixing percentage of total recurrence. 
A recurrence is marked if the distance between two vectors is below a threshold. We create a two-dimensional matrix, known as the recurrence matrix,
\begin{eqnarray}\label{eq1}
 \centering
    R_{i,j} = \Theta(\epsilon - \|X_{i}-X_{j}\|) \quad i,j = 1,2,...,n -(d-1)\tau
 \end{eqnarray}
  where $\Theta$ is the Heaviside step function. $\|X_{i}-X_{j}\|$ is the Euclidean distance between pairs of points, $i$ and $j$, in the reconstructed phase space. $\epsilon$ is a threshold to define recurrences in the reconstructed phase space. Here, $\epsilon$, is selected as 30 \% of the size of the attractor in the phase space. 
 Each element of the $R_{i,j}$ matrix corresponds to a pair of data points in the time series. If the distance between the pair of data points is below the threshold, they are marked as recurred and the corresponding matrix element is set to 1. Otherwise, it is set to 0. In the recurrence plot, recurrent points are represented as black or colored dots, while non-recurrent points are represented as white or empty spaces.

We can quantify recurrence patterns using different measures. Determinism is one measure that relates to the presence of regular or predictable patterns in the data. A signal or time series is considered more deterministic if it exhibits a high degree of order, coherence, or predictability.
% , indicating that its future behavior can be inferred from its past or current state. Conversely, a signal is considered less deterministic if it demonstrates randomness and a lack of predictability. Determinism quantifies the diagonal structures in the recurrence plot.
We count the number of diagonal lines in the recurrence plot that exceed a certain length $(l_{min})$. These diagonal lines represent consecutive recurrent states. The determinism (DET) is then computed as the ratio of the number of diagonal lines exceeding $l_{min}$ to the total number of recurrent points in the recurrence plot.
\begin{eqnarray}\label{eq1}
 \centering
    DET = \frac{\sum\limits_{l=l_{min}}^{N} lP(l)}{\sum\limits_{l=1}^{N} lP(l)}.
     \end{eqnarray}
The probability distribution of diagonal lines having length $l$ is denoted as $P(l)$. We use $l_{min} = 2$.

\section*{Appendix}
\subsection*{Evolution of the amplitude spectrum}

\begin{figure*}[h]
\begin{center}
\includegraphics[scale =0.36]{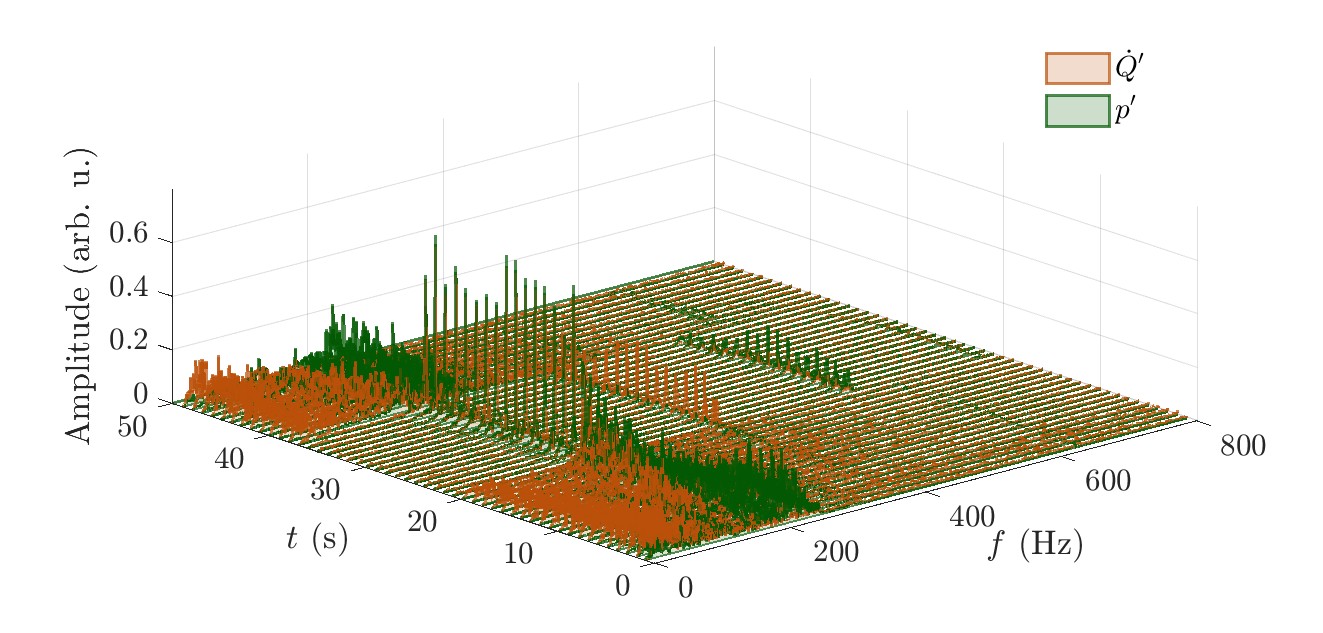}
\caption{\textbf{} Waterfall diagram showing the evolution of the dominant frequency of global heat release rate and acoustic pressure signal. }
\label{fig8}
\end{center}
\end{figure*}

Figure~\ref{fig8} shows the evolution of the amplitude spectrum of the global heat release rate and the acoustic pressure signals. It is evident that during the state of combustion noise, both the signals have a relatively broad distribution of frequencies with small peaks. However, during explosive synchronization, a sudden and abrupt transition to a sharp peak of frequency 190 Hz, with its harmonics, can be observed. As the system goes back to combustion noise, this common dominant peak disappears. The explosive nature of synchronization is evident from the sudden locking of dominant frequencies between the subsystems.

\subsection*{Robustness of the complex network}

\begin{figure*}[h!]
\begin{center}
\includegraphics[scale =0.38]{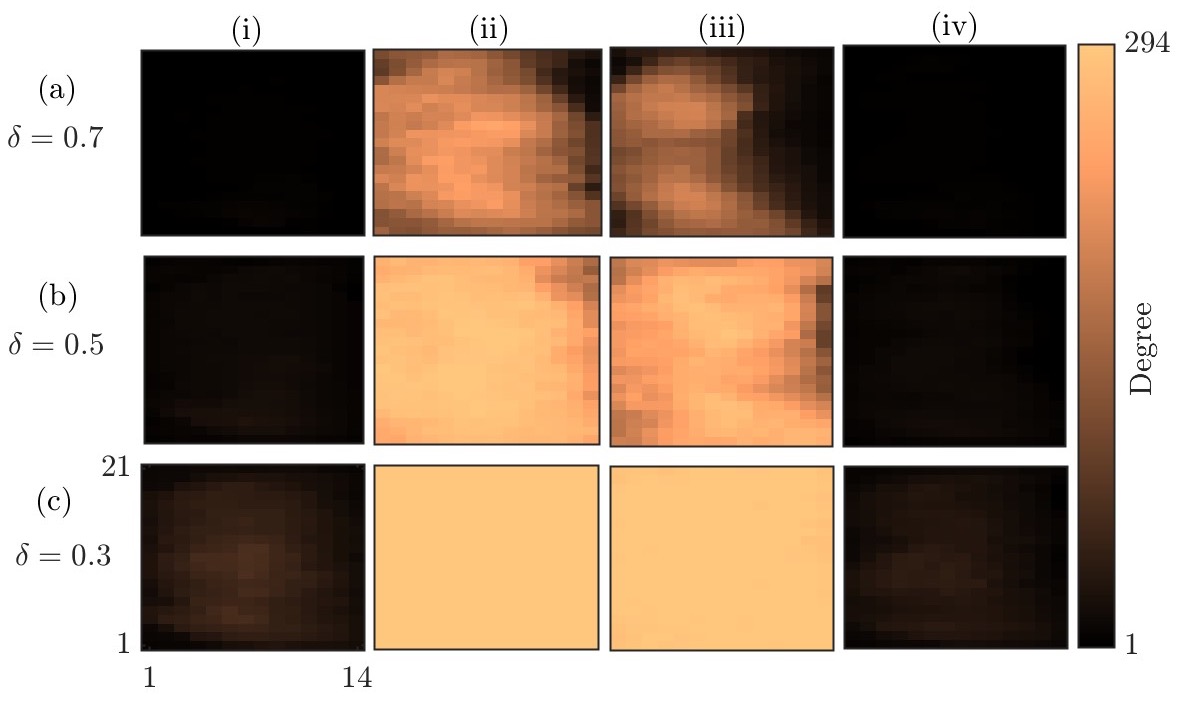}
\caption{\textbf{} Evolution of degree of the network for windows corresponding to $(i)$  $\mu _{yy}= 0.8$, $(ii)$ $\mu _{yy}= 0.9$, $(iii)$ $\mu _{yy}= 0$ and $(iv)$ $\mu _{yy}= -0.1$, computed with a threshold value of (a) $\delta = 0.7$ (b) $\delta = 0.5$ and (c) $\delta = 0.3$. The explosive nature of the synchronization is still preserved for all these thresholds. }
\label{fig9}
\end{center}
\end{figure*}

We construct complex networks from every 1 s window of time series and study the evolution of the network in time. We calculate the cross-correlation between each heat release rate oscillator and the acoustic field, and we use a threshold value of correlation to define the connections. This correlation threshold demarcates the synchronized and desynchronized dynamics. Figure \ref{fig9} shows the spatial distribution of degree (number connections of the oscillators) for different thresholds at four different instances. We can see that the network is robust to changes in this threshold within a range from 0.3 to 0.7, and it captures the explosive synchronization. 

\bibliography{ref}

\bibliographystyle{ScienceAdvances}

\section*{Acknowledgments}
We acknowledge P. R. Midhun, S. Anand, S. Thilagaraj, G. Sudha and M. Ragunathan for their help during the experiments. We thank the Science and Engineering Research Board (SERB) of the Department of Science and Technology for funding (Grant no: CRG/2020/003051). 
\section*{Author contributions}
R.I.S. conceived the idea. R.I.S. and I.P. designed the study. I.P. conducted the experiments. A.J. and I.P. performed data analysis and modeling. All authors discussed the results and participated in writing the manuscript.
\section*{Competing interests} The authors declare that they have no competing interests.
\section*{Data availability}
All the data presented in this paper are available from the corresponding author on reasonable request.

%Here you should list the contents of your Supplementary Materials -- below is an example. 
%You should include a list of Supplementary figures, Tables, and any references that appear only in the SM. 
%Note that the reference numbering continues from the main text to the SM.
% In the example below, Refs. 4-10 were cited only in the SM.     
% For your review copy (i.e., the file you initially send in for
% evaluation), you can use the {figure} environment and the
% \includegraphics command to stream your figures into the text, placing
% all figures at the end.  For the final, revised manuscript for
% acceptance and production, however, PostScript or other graphics
% should not be streamed into your compliled file.  Instead, set
% captions as simple paragraphs (with a \noindent tag), setting them
% off from the rest of the text with a \clearpage as shown  below, and
% submit figures as separate files according to the Art Department's
% instructions.
\end{document}